\documentclass[twocolumn,showpacs,preprintnumbers,amsmath,amssymb,prl]{revtex4-1}
\usepackage{epsfig} 
\usepackage{dcolumn}
\usepackage{bm}

\begin{document}

 
\title{Electric dipole matrix elements for the $6p\ ^2P_J \rightarrow 7s\ ^2S_{1/2}$ transition in atomic cesium}
\author{George Toh$^{1,2}$, Amy Damitz$^{2,3}$, Nathan Glotzbach$^{1,3}$, Jonah Quirk$^{3,4}$, I. C. Stevenson$^{1,2}$, J. Choi$^{1,2}$, M. S. Safronova$^{5,6}$, and D. S. Elliott$^{1,2,3}$}
\affiliation{%
   $^1$School of Electrical and Computer Engineering, Purdue University, West Lafayette, Indiana 47907, USA\\
   $^2$Purdue Quantum Center, Purdue University, West Lafayette, Indiana 47907, USA\\
   $^3$Department of Physics and Astronomy, Purdue University, West Lafayette, Indiana 47907, USA\\ 
   $^4$Pott College of Science, Engineering and Education, University of Southern Indiana, Evansville, Indiana 47712, USA \\
   $^5$Department of Physics and Astronomy, University of Delaware, Newark, Delaware 19716, USA \\
   $^6$Joint Quantum Institute, National Institute of Standards and Technology and the University of Maryland, College Park, Maryland 20742, USA }

\date{\today}

\begin{abstract}
We report a measurement of the ratio of electric dipole transition matrix elements of cesium for the $6p\,^2P_{1/2} \rightarrow 7s\,^2S_{1/2}$ and $6p\,^2P_{3/2} \rightarrow 7s\,^2S_{1/2}$ transitions.  We determine this ratio of matrix elements through comparisons of two-color, two-photon excitation rates of the $7s\,^2S_{1/2}$ state using laser beams with polarizations parallel to one another vs.\ perpendicular to one another.  Our result of $R \equiv \langle 7s\ ^2S_{1/2} || r || 6p\ ^2P_{3/2} \rangle / \langle 7s\ ^2S_{1/2} || r || 6p\ ^2P_{1/2} \rangle = 1.5272 \ (17)$ is in excellent agreement with a theoretical prediction of  $R=1.5270 \ (27)$. Moreover, the accuracy of the experimental ratio is sufficiently high to differentiate  between various theoretical approaches.  To our knowledge, there are no prior experimental measurements of $R$.  Combined with our recent determination of the lifetime of the $7s\,^2S_{1/2}$ state, we determine reduced matrix elements for these two transitions, $\langle 7s\ ^2S_{1/2} || r || 6p\ ^2P_{3/2} \rangle = -6.489 \ (5) \ a_0$ and $\langle 7s\ ^2S_{1/2} || r || 6p\ ^2P_{1/2} \rangle  = -4.249 \ (4) \ a_0$.  These matrix elements are also in excellent agreement with theoretical calculations. These measurements improve knowledge of Cs properties needed for parity violation studies and provide benchmarks for tests of high-precision theory.

\end{abstract}

                             
 \maketitle

\section{Introduction}

Precision values of atomic transition matrix  elements are needed for the determination of polarizabilities, light shifts and magic wavelengths for state-insensitive laser cooling, trapping, and atom manipulation \cite{SafronovaSC16,CooCovMad18}; long-range interaction $C_6$ and $C_8$ coefficients \cite{PorSafDer14}; blackbody radiation shifts \cite{NicCamHut15} and other systematic clock uncertainties \cite{LudBoyYe15}.
As a result, there is a critical need for benchmark measurements and calculations of electric-dipole and other transition matrix elements for various searches for physics
beyond the standard model of elementary particles \cite{SafBudDem18},  further improvement of current  atomic clocks \cite{NicCamHut15,HunSanLip16,McGZhaFas18} and development of novel frequency standards \cite{KozSafCre18}, study of degenerate quantum gases \cite{PagManCap15}  and quantum simulation \cite{ZhaBisBro14},  suppression of decoherence in quantum information processing \cite{ZhaRobSaf11,GolNorKol15}, etc. Most of these applications involve alkali-metal and alkaline-earth metal atoms and singly charged ions with similar electronic structure. Therefore, providing high-precision benchmark values for these systems and testing high-precision theory \cite{SafJoh08,PorBelDer09,TupKozsaf16} used for these applications is particularly important. There are particularly few high-precision (better than 0.5\%) benchmarks for the transitions between the excited states, which is the subject of this paper.

Laboratory determinations of the reduced electric dipole (E1) matrix elements of atomic cesium between the lowest $ns\ ^2S_{1/2}$ and $mp\ ^2P_{J}$ states, where $J=$1/2 or 3/2 is the electronic angular momentum of the state, are critical for calculations~\cite{PorBelDer09,PorsevBD10} of the parity nonconserving amplitude of the $6s\ ^2S_{1/2} \rightarrow 7s\ ^2S_{1/2}$ transition in cesium, as well as for precise calculation of the scalar and vector polarizability for this same transition~\cite{BlundellSJ92,SafronovaJD99,VasilyevSSB02}.
Atomic parity violation studies are uniquely sensitive to some dark matter candidates \cite{DavLeeMar14} and allow the study of hadronic parity violation in heavy nuclei \cite{HaxWie01}, not accessible by other experiments.
Most Cs experimental measurements focused on determinations of the 
$\langle 6s\ ^2S_{1/2}|| r || 6p\ ^2P_{J} \rangle$
matrix elements for the electric-dipole transitions from the ground state, which were measured through a number of means, including time-resolved fluorescence~\cite{YoungHSPTWL94,PattersonSEGBSK15}, absorption~\cite{RafacTLB99}, ground state polarizability~\cite{AminiG03,GregoireHHTC15}, and photoassociation spectroscopy~\cite{DereviankoP02,BouloufaCD07, ZhangMWWXJ13}, with good agreement between these independent results.  The weighted average of these measurements results in dipole moments with a precision of $\sim$0.035\%.  The $\langle 7s\ ^2S_{1/2}|| r || 7p\ ^2P_{J} \rangle$ moments were determined through Stark shift measurements~\cite{BennettRW99} of the $ 7s\ ^2S_{1/2}$ state, combined with theoretical results~\cite{SafronovaJD99} for the ratio $\langle 7s\ ^2S_{1/2}|| r || 7p\ ^2P_{3/2}   \rangle / \langle 7s\ ^2S_{1/2}|| r || 7p\ ^2P_{1/2}   \rangle$.  The precision of these moments is also very good, $\sim$0.15\%.
There are several measurements~\cite{VasilyevSSB02,antypas7p2013,Borvak14,DamitzTPTE18a} of the $\langle 7p\ ^2P_{J} || r || 6s\ ^2S_{1/2}  \rangle$ moments, with some significant differences among them. The precision of the most recent \cite{DamitzTPTE18a} is 0.12-0.14\%.
Finally, we recently reported~\cite{TohJGQSCWE18} a precise measurement of the lifetime of the cesium $7s\, ^2S_{1/2}$ state, $\tau_{7s} = 48.28 \ (7)$ ns, to a precision of 0.14\%.  Since this state spontaneously decays through two states (the $6p\, ^2P_{1/2}$ and the $6p\, ^2P_{3/2}$ state), the lifetime measurement by itself is not sufficient to determine the individual matrix elements for these two transitions. In this work, we present our determination of the ratio
\begin{equation}\label{eq:Rdefinition}
    R \equiv \frac{\langle 7s\ ^2S_{1/2} || r || 6p\ ^2P_{3/2} \rangle}{\langle 7s\ ^2S_{1/2} || r || 6p\ ^2P_{1/2} \rangle}
\end{equation}
based upon measurements of the influence of laser polarization on the two-photon $6s\ ^2S_{1/2} \rightarrow 7s\ ^2S_{1/2}$ transition rate.  This technique has been used previously~\cite{SieradzanHS04} to measure the branching ratio for spontaneous decay of the $8s \ ^2S_{1/2}$ state in cesium. Our result for $R$
 is in excellent agreement with a theoretical prediction and the accuracy of the experimental ratio is sufficiently high to differentiate  between various theoretical approaches. 
To our knowledge, there are no prior experimental determinations of this ratio.  We use the result of the current measurement, together with the lifetime measurement~\cite{TohJGQSCWE18}, to report E1 matrix elements $\langle 7s\ ^2S_{1/2} || r || 6p\ ^2P_{3/2} \rangle$ and $ \langle 7s\ ^2S_{1/2} || r || 6p\ ^2P_{1/2} \rangle$ with an uncertainty of 0.1\%.  These results are also in very good agreement with a number of prior theoretical calculations of these moments~\cite{DzubaFKS89,BlundellJS91,BlundellSJ92,SafronovaJD99,DzubaFG01,PorsevBD10,SafronovaSC16}.


\begin{figure} [t!]
	  \includegraphics[width=4cm]{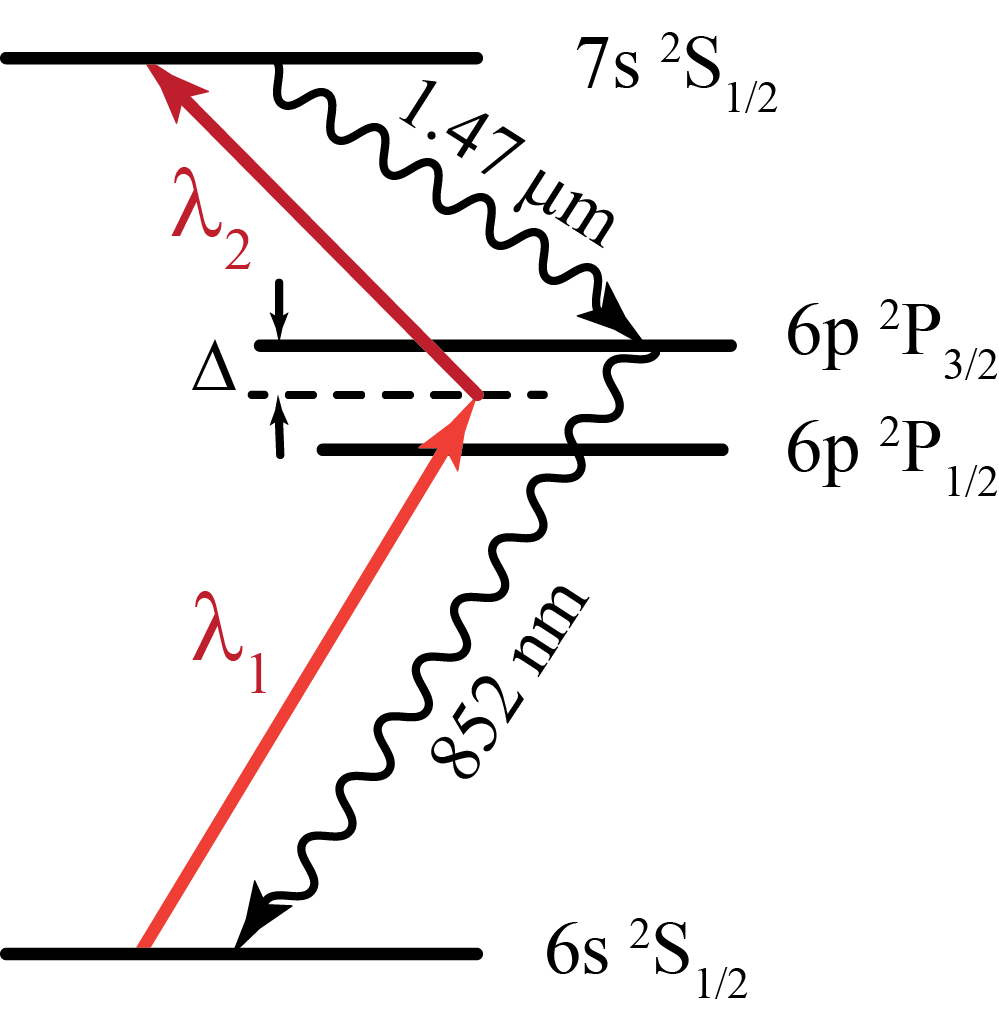}\\
	  \caption{Energy level diagram of atomic cesium, showing the states relevant to this measurement. Atoms are excited from the $6s\, ^2S_{1/2}$ ground state to the $7s\, ^2S_{1/2}$ excited state by two-color, two-photon excitation. We collect fluorescence photons at 852 nm from the second step of the spontaneous decay of atoms from the 7s state to the ground state by way of the $6p\,^2P_{3/2}$ state. }
	  \label{fig:EnergyLevel}
\end{figure}

\section{Theory}
For this determination of $R$, we carry out a series of measurements of the two-color, two-photon $6s\ ^2S_{1/2} \rightarrow 7s\ ^2S_{1/2}$ absorption rate.  
The first laser for this excitation is tuned to a frequency between the resonant frequency of the $6s\ ^2S_{1/2} \rightarrow 6p\ ^2P_{3/2}$ transition (D$_2$) and that of the $6s\ ^2S_{1/2} \rightarrow 6p\ ^2P_{1/2}$ transition (D$_1$), as illustrated in Fig.~\ref{fig:EnergyLevel}.  The detuning of this laser from the D$_2$ line frequency is labeled $\Delta$.  The frequency of the second laser is tuned to complete the two-photon transition to the $7s\ ^2S_{1/2}$ state.  
For determination of the ratio of moments $R$, we compare the two-photon signal strength using two laser polarization states as a function of the detuning $\Delta$ from the intermediate resonance.  In both cases, the two lasers are linearly polarized, with the relative polarizations either parallel or perpendicular to one another.  

We can quantitatively understand the dependence of the two-photon transition rate on polarization by examining the two-photon transition rate expressed through the Fermi golden rule 
\begin{equation}
  S = \frac{2 \pi}{\hbar} | A_{\rm 2P} |^2 \rho_{\rm 7s}(E),
\end{equation}
where $\rho_{\rm 7s}(E)$ is the final state energy density and $A_{\rm 2P}$ is the transition amplitude as determined in lowest-order perturbative expression
\begin{eqnarray}
  A_{\rm 2P} &=& \sum_{n,j} \left\{ \frac{\langle 7s_{1/2} | \hat{\mbox{\boldmath $\varepsilon$}}_1 E_1 \cdot e \mathbf{r} | np_j \rangle \langle np_j | \hat{\mbox{\boldmath $\varepsilon$}}_2 E_2 \cdot e \mathbf{r} | 6s_{1/2} \rangle  }{\omega_2 - \omega_{np_j} -  i\Gamma_{np_j}/2} \right. \nonumber \\
  & & + \left. \frac{\langle 7s_{1/2} | \hat{\mbox{\boldmath $\varepsilon$}}_2 E_2 \cdot e \mathbf{r} | np_j \rangle \langle np_j | \hat{\mbox{\boldmath $\varepsilon$}}_1 E_1 \cdot e \mathbf{r} | 6s_{1/2} \rangle  }{\omega_1  - \omega_{np_j} - i\Gamma_{np_j}/2} \right\}. \nonumber
\end{eqnarray}
In this expression, we have abbreviated the state notation $|m \ell \  ^2L_J \rangle $ by the single active electron $|m \ell_j \rangle $.  The polarization, amplitude, and frequency of the optical fields are $\hat{\mbox{\boldmath $\varepsilon$}}_1$, $E_1$, and $\omega_1$ for the first laser beam, of wavelength $\sim 860$ nm, and $\hat{\mbox{\boldmath $\varepsilon$}}_2$, $E_2$, and $\omega_2$ for the second, of wavelength $\sim 1.45 \ \mu$m. $ \mathbf{r}$ is the spatial coordinate of the electron, and $\omega_{np_j}$ and $\Gamma_{np_j}/2$ are the transition frequency from the ground state and the radiative linewidth of the intermediate states $np_j$.  The detunings that we use in the measurements are always much larger than the linewidth $\Gamma_{np_j}/2$, and we omit the linewidth term from our analysis.

The ground state of the cesium atom is split by the hyperfine interaction into two components, $F=3$ and $F=4$, separated by 9.1926 GHz.  $F$ is the total angular momentum (electronic $J=1/2$ plus nuclear $I=7/2$) of the state.  Similarly, the final $7s\ ^2S_{1/2}$ has two hyperfine components, also $F=3$ and $F=4$, with a splitting of $2.183 $ GHz~\cite{GilbertWW83, YangWYW16}. The transition moment for a particular hyperfine component is given through the Wigner-Eckart theorem (See, for example, Ref.~\cite{zare1988angular}, page 192.) as
\begin{eqnarray}
  \langle \gamma J I F &m_F& | r_q | \gamma^{\prime} J^{\prime} I^{\prime} F^{\prime} m_F^{\prime} \rangle = (-1)^{F - m_F} \nonumber \\ 
  & \times &  \left( \begin{array}{c c c}
  F & 1 & F^{\prime} \\ -m_F & q & m_F^{\prime}  \end{array} \right)  \langle \gamma J I F || r || \gamma^{\prime} J^{\prime} I^{\prime} F^{\prime}  \rangle, \nonumber
\end{eqnarray}
where $m_F$ is the projection of the total angular momentum onto the quantization axis, and $\gamma$ represents all other quantum numbers, shows how the moments vary with projection quantum number $m_F$.  The array inside the smooth parentheses is the Wigner $3j$ symbol.  Since $r$ acts only on the electronic angular momentum, but not $I$, we can further reduce this using (See, for example, Ref.~\cite{zare1988angular}, page 195.)
\begin{eqnarray}
  \langle \gamma &J& I F | r | \gamma^{\prime} J^{\prime} I^{\prime} F^{\prime}  \rangle  = \delta_{I I^{\prime}}   (-1)^{J + I + F^{\prime} + 1}  \nonumber \\ 
  & \times & \left[ \left( 2F^{\prime} + 1 \right) \left( 2F + 1 \right) \right]^{1/2}  \left\{ \begin{array}{c c c}
  J & F & I \\ F^{\prime} & J^{\prime} & 1  \end{array} \right\}  \langle \gamma J || r || \gamma^{\prime} J^{\prime}  \rangle. \nonumber
\end{eqnarray}
The array inside the brackets is the Wigner $6j$ symbol. 
These relations allow calculation of all of the moments relevant for the two-photon absorption process.  Since the initial population is equally distributed over the sixteen hyperfine components of the ground state, and we spectrally resolve the hyperfine states $F$ of the initial $6s\ ^2S_{1/2}$ and final $7s\ ^2S_{1/2}$ state, we average the moments over initial state components $m_F$ after squaring, and we sum over final states $m_F^{\prime}$ to obtain a two-photon signal strength $S$ 
\begin{equation}\label{eq:Spar44}
  S_{\parallel, 4 \rightarrow 4} = \frac{2 \pi}{\hbar^2} \frac{9}{16}  |\tilde{\alpha}|^2 E_1^2 E_2^2,
\end{equation}
where
\begin{eqnarray}\label{eq:alpha}
  \tilde{\alpha} &=& \frac{e^2}{6} \sum_{n}  \left[ \rule{0in}{0.2in}  \langle 7s_{1/2} || r || np_{1/2}  \rangle \langle np_{1/2} || r || 6s_{1/2} \rangle \right. \nonumber \\
  & & \hspace{0.5in} \times \left\{ \frac{1}{\omega_2 -\omega_{np_{1/2}} } + \frac{1}{\omega_1 - \omega_{np_{1/2}} } \right\}   \\
  & & \hspace{0.3in}  - \rule{0in}{0.2in}  \langle 7s_{1/2} || r || np_{3/2}  \rangle \langle np_{3/2} || r || 6s_{1/2} \rangle  \nonumber \\
  & & \hspace{0.5in} \left. \times \left\{ \frac{1}{\omega_2 - \omega_{np_{3/2}} } + \frac{1}{\omega_1 - \omega_{np_{3/2}} } \right\} \right]   \nonumber 
\end{eqnarray}
for parallel polarization on the $6s \ ^2S_{1/2}, \ F = 4 \rightarrow 7s \ ^2S_{1/2}, \  F=4$ component ($4 \rightarrow 4$).  For the perpendicular polarization case, the two-photon $ 4 \rightarrow 4$ signal is
\begin{equation}\label{eq:Sperp44}
  S_{\perp, 4 \rightarrow 4} = \frac{2 \pi}{\hbar^2} \frac{15}{64}  |\tilde{\beta}|^2 E_1^2 E_2^2,
\end{equation}
where
\begin{eqnarray}\label{eq:beta}
 \tilde{\beta} &=& \frac{e^2 }{6} \sum_{n}  \left[ \rule{0in}{0.2in}  \langle 7s_{1/2} || r || np_{1/2}  \rangle \langle np_{1/2} || r || 6s_{1/2} \rangle \right. \nonumber  \\
  & & \hspace{0.5in} \times \left\{ \frac{1}{\omega_2 - \omega_{np_{1/2}} } - \frac{1}{\omega_1 - \omega_{np_{1/2}} } \right\}  \\
  & & \hspace{0.3in}  + \rule{0in}{0.2in}  \frac{1}{2} \langle 7s_{1/2} || r || np_{3/2}  \rangle \langle np_{3/2} || r || 6s_{1/2} \rangle  \nonumber \\
  & & \hspace{0.5in} \left. \times \left\{ \frac{1}{\omega_2 - \omega_{np_{3/2}} } - \frac{1}{\omega_1 -  \omega_{np_{3/2}} } \right\} \right] .  \nonumber 
\end{eqnarray}
The ratio of these two linestrengths is
\begin{equation}\label{eq:ratio44}
  \left( \frac{S_{\parallel}}{S_{\perp} }\right)_{ 4 \rightarrow 4} = \frac{12}{5} \ \frac{|\tilde{\alpha}|^2}{|\tilde{\beta}|^2} .
\end{equation}

Similarly, on the $6s \ ^2S_{1/2}, \ F = 3 \rightarrow 7s \ ^2S_{1/2}, \  F=3$ component ($3 \rightarrow 3$) component, the linestrengths are
\begin{equation}\label{eq:Spar33}
  S_{\parallel, 3 \rightarrow 3} = \frac{2 \pi}{\hbar^2} \frac{7}{16}  |\tilde{\alpha}|^2 E_1^2 E_2^2
\end{equation}
for parallel polarization, and 
\begin{equation}\label{eq:Sperp33}
  S_{\perp, 3 \rightarrow 3} = \frac{2 \pi}{\hbar^2} \frac{7}{64}  |\tilde{\beta}|^2 E_1^2 E_2^2
\end{equation}
for perpendicular polarization.
The ratio of these two linestrengths is
\begin{equation}\label{eq:ratio33}
  \left( \frac{S_{\parallel}}{S_{\perp} }\right)_{ 3 \rightarrow 3} = 4 \  \frac{|\tilde{\alpha}|^2}{|\tilde{\beta}|^2} .
\end{equation}

Transitions on the $F = 3 \rightarrow F=4$ and the $F = 4 \rightarrow F=3$ components are also permitted for the perpendicular polarization case, but our spectral resolution is sufficient to avoid these components, and we do not consider them further.

Due largely to the small magnitude of the detuning of the first laser from the D$_1$ and D$_2$ lines, the dominant contributions to the two-photon moments in Eqs.~(\ref{eq:alpha}) and (\ref{eq:beta}) are from the $6p\ ^2P_{3/2}$ and $6p\ ^2P_{1/2}$ states.  Similar to the approach of Ref.~\cite{SieradzanHS04}, we factor out the product of elements $\langle 7s_{1/2} || r || 6p_{1/2}  \rangle \langle 6p_{1/2} || r || 6s_{1/2} \rangle $, allowing us to show explicitly the dependence of $\tilde{\alpha}$ and $\tilde{\beta}$ on the ratio of dipole elements $R$: 
\begin{eqnarray}\label{eq:alpha_R}
  \tilde{\alpha} &=&  K  \left[ \frac{R \ (-R^{\prime})}{\omega_1 -\omega_{6p_{3/2}} } + \frac{1}{\omega_1 - \omega_{6p_{1/2}} } \right. \\
  & & \hspace{0.5in} \left. + \frac{R \ (-R^{\prime})}{\omega_2 - \omega_{6p_{3/2}} } + \frac{1}{\omega_2 - \omega_{6p_{1/2}} } + P \right]   \nonumber
\end{eqnarray}
and
\begin{eqnarray}\label{eq:beta_R}
 \tilde{\beta} &=& K  \left[ \frac{R \ (-R^{\prime}/2)}{\omega_1 -\omega_{6p_{3/2}} } - \frac{1}{\omega_1 - \omega_{6p_{1/2}} } \right. \\
  & & \hspace{0.5in} \left. - \frac{R \ (-R^{\prime}/2)}{\omega_2 - \omega_{6p_{3/2}} } + \frac{1}{\omega_2 - \omega_{6p_{1/2}} } + Q \right] ,  \nonumber  
\end{eqnarray}
where 
\begin{displaymath}\label{eq:constantK}
  K = \frac{e^2}{6} \langle 7s_{1/2} || r || 6p_{1/2}  \rangle \langle 6p_{1/2} || r || 6s_{1/2} \rangle 
\end{displaymath}
and 
\begin{eqnarray}
R^{\prime} = \frac{ \langle 6p_{3/2} \| r \| 6s_{1/2} \rangle}{ \langle 6p_{1/2} \| r \| 6s_{1/2} \rangle}. 
	\label{eqn:Rprime}
\end{eqnarray}
We use the weighted mean of the measured~\cite{YoungHSPTWL94, RafacTLB99, DereviankoP02, AminiG03, BouloufaCD07, ZhangMWWXJ13,PattersonSEGBSK15,GregoireHHTC15} values of transition moments $\langle 6p_{1/2} || r || 6s_{1/2}  \rangle = 4.5057 \ (16) \ a_0$ and $\langle 6p_{3/2} || r || 6s_{1/2} \rangle = -6.3398 \ (22) \ a_0$ to determine 
\begin{displaymath}
  R^{\prime} = -1.40706 \ (70).
\end{displaymath}

$P$ accounts for the rather minor contributions of the high $n$ states (that is, $n>6$) to the parallel polarization signal
\begin{equation}
P =    \sum_{n > 6,j,k} \frac{(-1)^{j-1/2} M_{nj} }{\omega_k - \omega_{np_j}} ,
\label{eqn:sumP}
\end{equation}
where $j = 1/2$ or $3/2$ and the index $k$ selects one of the two laser frequencies.  The term $Q$ performs the same role for the perpendicular polarization signal
\begin{equation}
Q =    \sum_{n > 6,j,k} \frac{(-1)^{k}}{j+1/2}  \ \frac{M_{nj} }{\omega_k - \omega_{np_j}} .
	\label{eqn:sumQ}
\end{equation}
In these expressions for $P$ and $Q$, the $M_{nj}$ are normalized products of dipole moments for $6s_{1/2} \rightarrow np_j \rightarrow 7s_{1/2}$, 
\begin{equation}
M_{nj} = \frac{ \langle 7s_{1/2} \| r \| np_{j} \rangle \langle np_{j} \| r \| 6s_{1/2} \rangle}{ \langle 7s_{1/2} \| r \| 6p_{1/2} \rangle \langle 6p_{1/2} \| r \| 6s_{1/2} \rangle}
	\label{eqn:ratioM}
\end{equation}



In the experiment, we measure the two-photon excitation signals $S_{\parallel}$ for parallel polarizations and $S_{\perp}$ for perpendicular polarizations over a wide range of detunings $\Delta$, and compute the ratio of these signals $S_{\parallel}/S_{\perp}$ to remove any dependence on laser power, beam size, collection efficiencies, detection sensitivities, and other experimental factors.  
In the following section, we discuss the experimental details of these measurements.

\section{Experimental details}

\begin{figure}
	  \includegraphics[width=8cm]{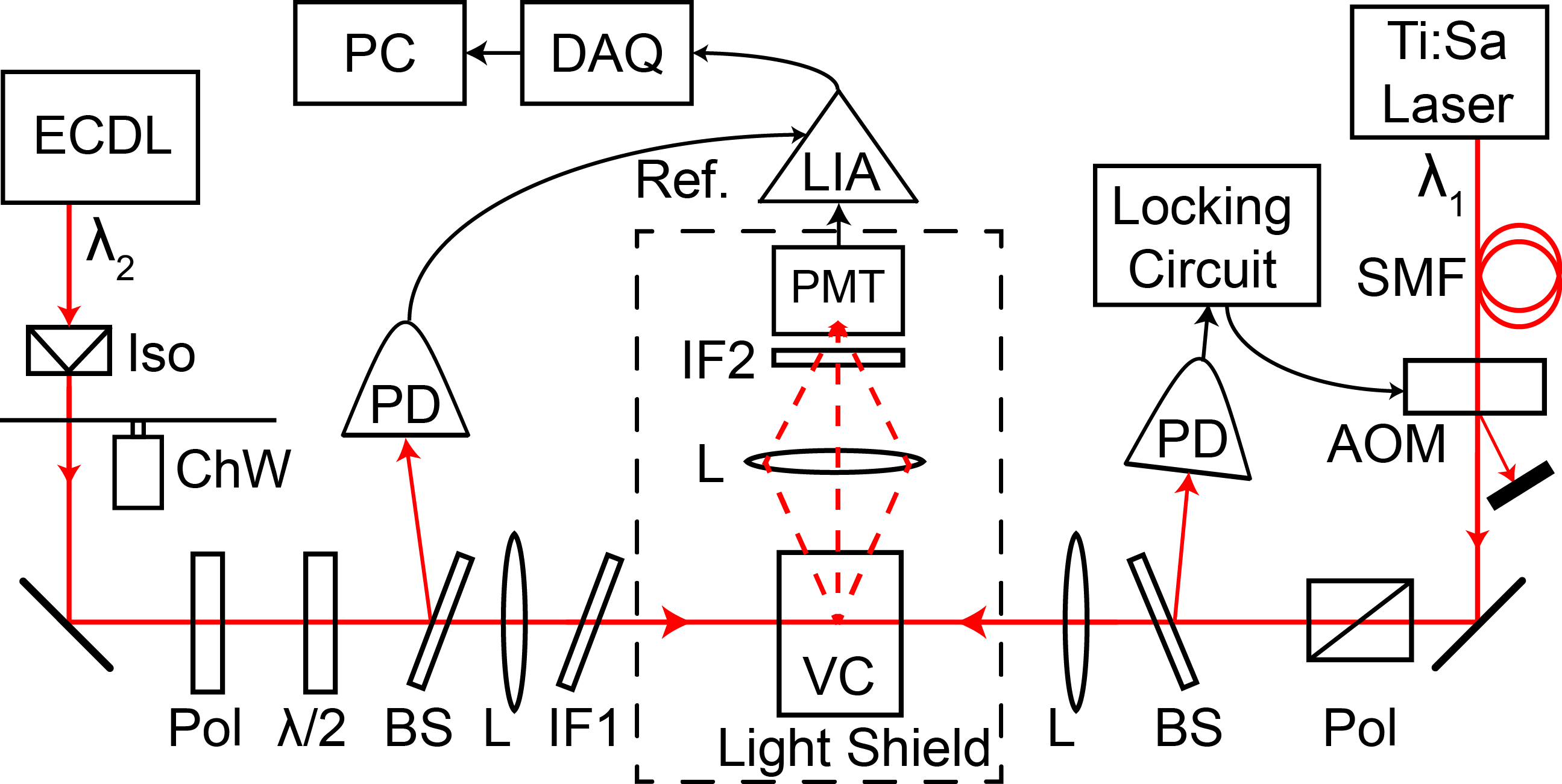}\\
	  \caption{Experimental setup for the measurement of the relative peak amplitudes with perpendicular and parallel laser polarizations. We keep the polarization of the Ti:Sapphire (Ti:Sa) laser beam constant and change the polarization of the external cavity diode laser (ECDL) beam. Other abbreviations in this figure are: (AOM) acousto-optic modulator; (BS) beam sampler; (ChW) beam chopper wheel; (DAQ) data acquisition system; (IF) interference filter; (Iso) optical isolator; (L) lens; (LIA) lock-in amplifier; (PC) personal computer; (PD) photodetector; (PMT) photomultiplier tube; (Pol) polarizer; (SMF) single-mode optical fiber, (VC) vapor cell, ($\lambda/2$) half-wave plate in a rotation stage.}
	  \label{fig:ExperimentalSetup}
\end{figure}

We use two narrowband cw lasers for these measurements. We show our experimental setup in Fig.~\ref{fig:ExperimentalSetup}. The first beam, whose wavelength $\lambda_1$ we vary in the range $855 - 870$ nm, is from a Ti:Sapphire laser, red-detuned from the Cs D$_2$ line at 852 nm. The Ti:Sapphire beam is sent over a single-mode optical fiber to the optical table where we conduct the experiment.
The second beam, at $\lambda_2 = 1415 - 1460$ nm, is blue-detuned from the Cs $6p_{3/2} \rightarrow 7s_{1/2}$ transition and is generated by a homemade external cavity diode laser (ECDL). The diode is a Toptica anti-reflection-coated laser diode. With the laser in a Littman configuration, we can coarsely tune this ECDL from $\sim 1400 - 1480$ nm without variation of the output beam direction. We measure the frequency of the ECDL beam with a calibrated Burleigh WA-1600 (Michelson interferometer type) wavemeter, with an accuracy of better than 0.1 GHz. Then we adjust the frequency of the Ti:Sapphire laser to place the two-photon resonance peak at the center of a 2.5 GHz scan, and ramp the frequency of the Ti:Sapphire laser at a rate of about 2.0 GHz/sec. 

After two-photon excitation of the $7s\ ^2S_{1/2}$ state, the atoms decay spontaneously to the ground state by way of the $6p\ ^2P_{3/2}$ or the $6p\ ^2P_{1/2}$ state.  
We detect the fluorescence light on the D$_2$ line at 852 nm as a measure of the excitation rate of the $7s\ ^2S_{1/2}$ state (see Fig. \ref{fig:EnergyLevel}).  We chose to collect this fluorescence line since the sensitivity of our photomultiplier tube (PMT, Hamamatsu R928) is greater at this wavelength than at the wavelengths of the other fluorescence lines.  
We chop the ECDL beam ($\sim 266$ Hz chopping rate) and amplify the PMT output with a lock-in amplifier to improve the signal-to-noise ratio of our detection system.  The output from the lock-in amplifier is read with a data acquisition (DAQ) system and recorded on the laboratory computer (PC).

The polarization purity of both laser beams passing through the vapor cell is critical for an accurate measurement.  We pass the Ti:Sapphire beam through a Glan-Taylor polarizer with extinction ratio $>$10,000:1. The ECDL beam is put through a nanoparticle linear film polarizer (extinction ratio $>$10,000:1), then through a zero-order half-wave plate (HWP) optimized for 1.48 $\mu$m.  We found that a polarizer after the half-wave plate 
could displace the beam, so we removed this element. To avoid introducing strain birefringence in any optics within the beam path after the polarizers (lens, beam sampler and wave-plate), we mounted these optics with soft plastic O-rings, or bonded them with flexible epoxy. (When the optics were mounted with hard epoxy and metal O-rings, we noticed a ten-fold reduction in laser extinction ratio.)  We suspected that the nanoparticle film polarizer was sensitive to the presence of the $\lambda_1$ beam, so we inserted a long-pass interference filter (IF1) in the beam to reflect the $\lambda_1$ beam after passing through the vapor cell.  The Ti:Sapphire laser beam passing through the vapor cell had a typical extinction ratio of a part in 10,000, while the extinction ratio of the second laser varied from a part in $10,000-200$, falling as we tuned away from the center frequency of the 1480 nm half-wave plate. We recorded the extinction ratio at every laser detuning to apply the proper correction to our data.

We weakly focus the two laser beams ($\lambda_1$ and {$\lambda_2$) with 15 cm focal length lenses through a cesium vapor cell (VC) in a counter-propagating configuration. 
The diameter of each beam in the vapor cell is $\sim$80 $\mu$m. 
The laser power passing through the vapor cell was $\sim$20 mW for the Ti:Sapphire beam and $\sim$5 mW for the ECDL beam, varying for each wavenumber measurement. We reduce the optical power for small detunings $\Delta$ in order to avoid saturating the transition.  
The cesium vapor cell is a fused silica cell with dimensions 70 x 10 x 10 mm$^3$. We place the vapor cell and PMT within an aluminum enclosure to reduce scattered light and to maintain a uniform cell temperature. We pass the laser beams close to the end of the cell near the PMT to minimize re-absorption of the fluorescence light, and image the interaction region with a lens of 1 inch focal length and 1 inch diameter. We place an interference filter (IF2) in front of the PMT, transmitting light at $850 \pm 5$ nm (Thorlabs FBH850-10), and also place a 6 mm $\times$ 2 mm spatial aperture in the image plane of the lens. These two filters reduce the light scattered by the entrance and exit faces of the cell into the PMT. 
We heated the cell with a cartridge heater to approximately $T \sim 140^\circ $C to attain sufficient cesium density for the measurement.  In the counter-propagating beam geometry, the Doppler width of the transition is 
\begin{equation}
\Delta \nu_D = \sqrt{\frac{8 k_B T \ln{2}}{m_{Cs}} } \ (\lambda_1^{-1} - \lambda_2^{-1}) \sim 170 \ {\rm MHz}, 
\end{equation}where $k_B$ is the Boltzmann constant and $m_{Cs}$ is the mass of the cesium atom.  This linewidth is much less than the hyperfine splitting of the $7s$ state, so the spectral lines that we measure are far removed from unwanted adjacent transitions.

We monitor the power of each laser beam by reflecting a small portion of the beams with Thorlabs beam samplers to photodetectors (PD). These beam samplers are wedged windows, AR-coated on one side and uncoated on the other. The power of the $\lambda_2$ beam transmitted by this window changes by $<$1\% (due to Fresnel reflection) when we rotate the polarization of this beam. Corrections we made to the data for these differences are discussed in the next section. In addition to monitoring the power of the ECDL during each data set, this PD produces the reference signal for the lock-in amplifier described earlier.  We use the Ti:Sapphire beam PD and an acousto-optic modulator (AOM) to stabilize the power of this beam. The closed-loop feedback circuit stabilizes the laser power against any fluctuations of the Ti:Sapphire laser as we ramp its frequency. 

We use Labview to record each of the fluorescence peaks and fit them to a Gaussian lineshape. 
In Fig.~\ref{fig:Strongpeak}, we show examples of the fluorescence peaks for individual scans at $\Delta/2\pi = 107.5$ cm$^{-1}$ for (a) parallel and (b) perpendicular polarizations.  The black points in this figure are the data, and the smooth red line is the result of the least-squares fit. 
\begin{figure}[b!]
      \includegraphics[width=8.5cm]{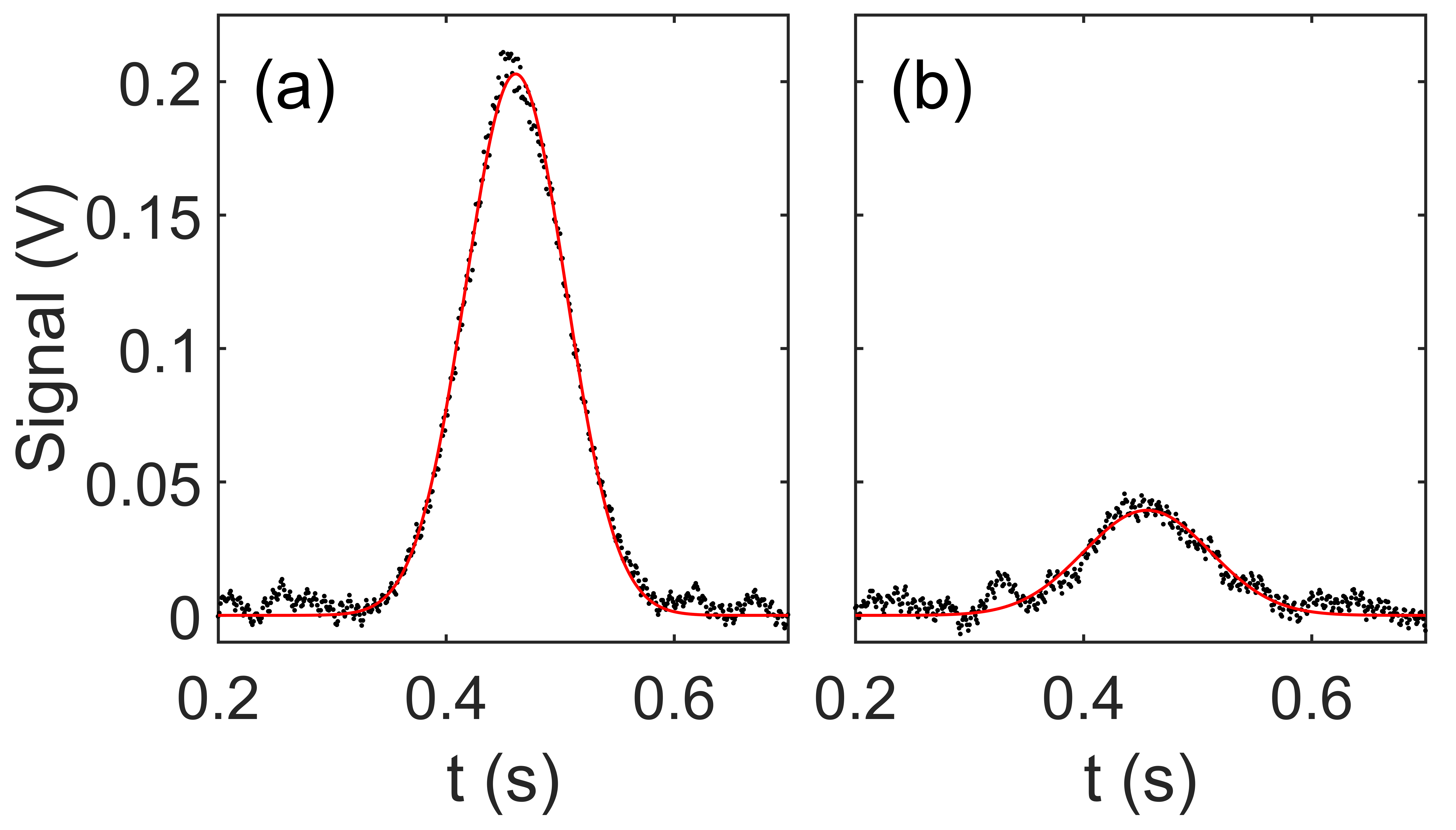}
	  \caption{Absorption spectra for (a) parallel polarization and (b) perpendicular polarization, at a detuning of $\Delta/2 \pi  = 107.5 $ cm$^{-1}$. The black dots are experimental data while the red curve is the least-squares Gaussian fit to the data.  With a laser frequency scan rate of $\sim$2.0 GHz/sec, the total frequency width of these plots is $\sim$1.0 GHz. The ratio of peak heights at this detuning is $S_{\parallel}/S_{\perp} \sim 5.35$.}
	  \label{fig:Strongpeak}
\end{figure}
In approximately two minutes we record thirty peaks, and determine the average and standard deviation of peak heights computed over the entire set.
We manually rotate the half-wave plate mounted in a rotation stage, changing the laser polarization of the ECDL beam between vertical (parallel to the polarization of the $\lambda_1$ beam) and horizontal (perpendicular).
We switch the polarization of the ECDL back and forth to acquire at least three measurements at each polarization. 
We then change the frequencies of the ECDL and Ti:Sapphire laser and repeat the process.

We calculate the ratio of line strengths $S_{\parallel}/S_{\perp}$ at a particular detuning by dividing the mean amplitude of the parallel peaks with the mean amplitude of the perpendicular peaks.
We plot the measured ratios $S_{\parallel}/S_{\perp}$ vs.\ detuning $\Delta / 2 \pi$ in Fig.~\ref{fig:ratiovsdetuning}.  
\begin{figure}
      \includegraphics[width=8cm]{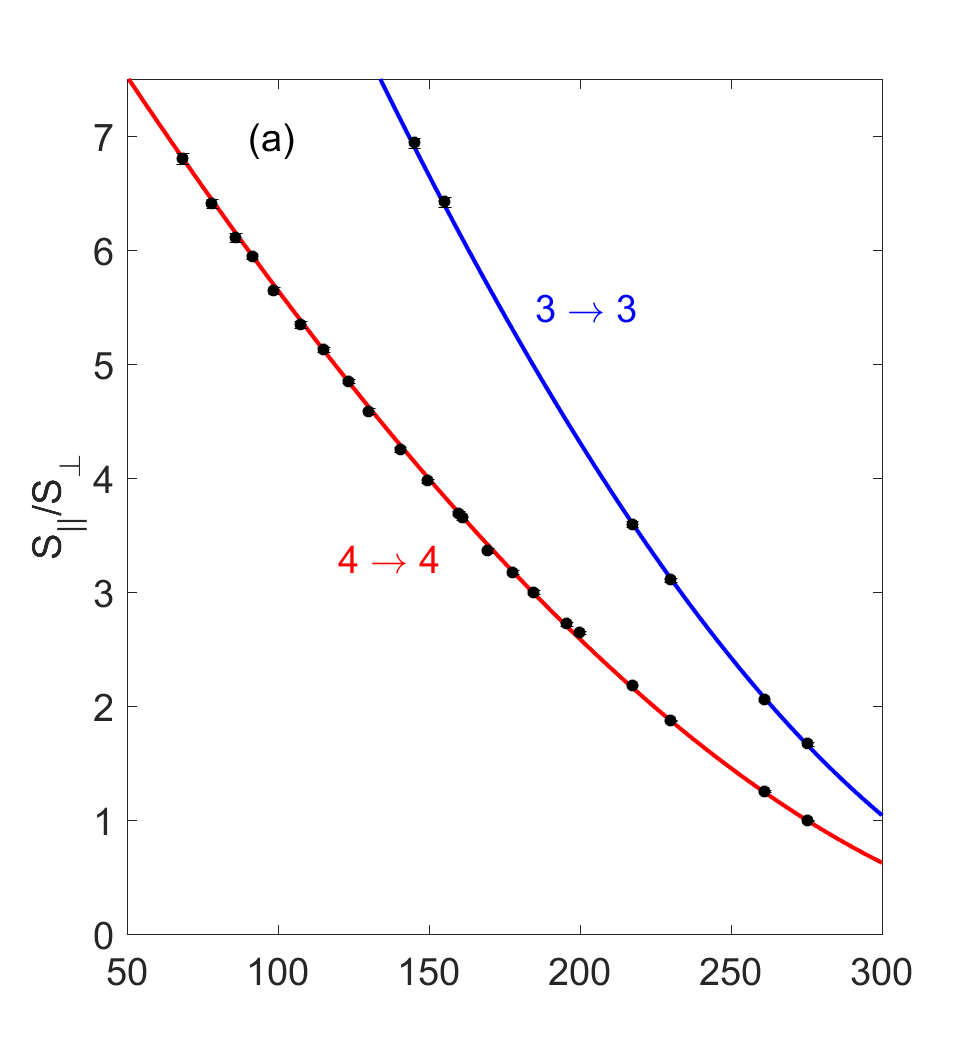} 
      \includegraphics[width=8cm]{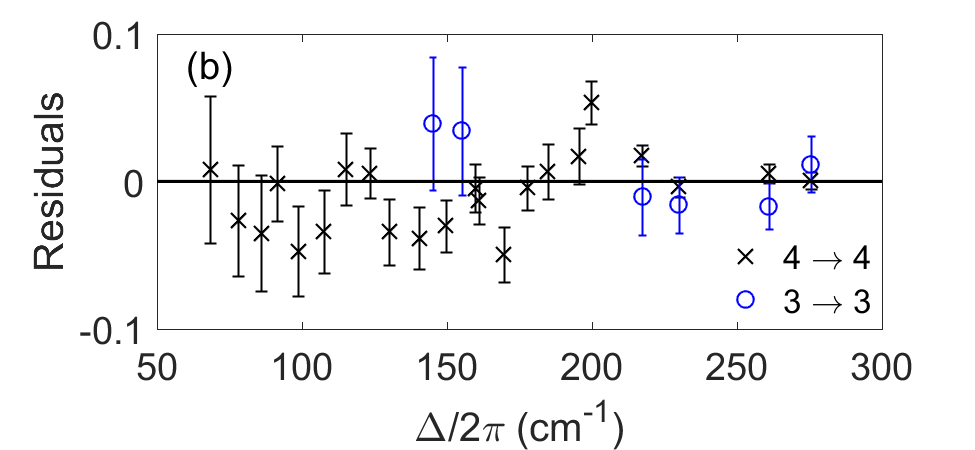} 
      \caption{(a) The ratio of peak heights $S_{\parallel}/S_{\perp}$ as a function of detuning $\Delta/2 \pi $. The lower (red) curve is for the $4 \rightarrow 4$ transition while the upper (blue) curve is for the $3 \rightarrow 3$ transition. The data points are the experimental data, with error bars showing $1 \sigma$ uncertainties. (In many cases, the uncertainties are smaller than the data point size.)  The smooth lines show the best fit plots of Eqs.~(\ref{eq:ratio44}, \ref{eq:ratio33}-\ref{eq:beta_R}), with $R = 1.5272$ the only adjustable parameter. (b) Residuals, showing the difference between data points and the fitted curve in (a). $4 \rightarrow 4$ residuals are shown with an $\times$, while $3 \rightarrow 3$ residuals are shown with an open circle ($\circ$). }
	  \label{fig:ratiovsdetuning}
\end{figure}
Each data point represents the experimental measurement, as described above.  The error bars show the $1 \sigma$ standard error.  
While we focused most of our attention on the $4 \rightarrow 4$ transition due to the smaller ratio $S_{\parallel}/S_{\perp}$ for this line, we did repeat the measurement in the accessible range of the $3 \rightarrow 3$ line to verify our results. We have plotted these points in Fig.~\ref{fig:ratiovsdetuning} as well. 
For the $4 \rightarrow 4$ transition we were able to collect data over a $>$200 cm$^{-1}$ range of $\Delta/2 \pi$, from 60 to 280 cm$^{-1}$.  We avoided detunings smaller than 60 cm$^{-1}$ for three reasons: The scattered light background is large in this region; the ratio $S_{\parallel}/S_{\perp}$ is less sensitive to $R$ at small detunings; and the peak height ratio is large at small detunings, making it difficult to simultaneously keep $S_{\parallel}$ below the saturation level (at the 0.1\% level) and $S_{\perp}$ sufficiently greater than the noise.

\section{Error analysis}
In addition to the statistical error, there are several other possible sources of error in performing this measurement.  
We summarize these effects and present estimates of their impact as a correction and uncertainty in $S_{\parallel}/S_{\perp}$ in Table~\ref{table:sourcesoferror}. We apply these corrections and expand the error bars to the individual $S_{\parallel}/S_{\perp}$ measurements before fitting the data.
\begin{table}[b!]
    \begin{tabular}{l|c|c}
        \hline
        \multicolumn{1}{c|}{Error}   		& \% Correction & \% Uncertainty  \\ \hline \hline
        Statistical							&				& $0.26 - 1.17$ \\
        Polarization purity					& $0.12 - 0.35$ & 0.05 \\
        Beam movement 						& 				& 0.01 \\
        Beam power change					& $0.1 - 1$		& $0.1 - 0.3$ \\
		HWP rotation precision				& 0.05			& 0.05 \\
        Magnetic field						& -0.1			& 0.1 \\
    \hline 
    \end{tabular}
    \caption{Sources of error and the correction applied to $S_{\parallel}/S_{\perp}$ and uncertainty for each.  We compute the uncertainty for each data point in Fig.~\ref{fig:ratiovsdetuning} as the quadrature sum of these contributions.}
    \label{table:sourcesoferror}
\end{table}

We previously discussed the polarization quality of the two laser beams, which varies with detuning $\Delta$ for the $\lambda_2$ beam. We monitored this carefully during the course of the measurements, and applied a correction to the ratio $S_{\parallel}/S_{\perp}$ to account for this. This correction was as large as 0.5\%, but typically $0.12 - 0.35 $\%. We estimate that the uncertainty in this correction is on the order of $0.05$\%.  

In addition, we must quantify the change in beam overlap and beam power as we rotate the half-wave plate in the $\lambda_2$ beam path. 
The beam displacement $\Delta x$ is smaller than we can measure in our laboratory, so we use the manufacturer's specification for the parallelism of the waveplate ($<5 \ \mu$rad) to estimate the beam displacement ($< 0.4 \ \mu$m) at the focus of the beam upon rotating the waveplate.  We have calculated that this introduces a fractional uncertainty of the measured ratio $S_{\parallel}/S_{\perp}$ of $(\Delta x / w)^2$, where $w$ is the beam radius.  This fractional uncertainty is less than 0.01\%.  
As we wrote in the previous section, the laser power of the $\lambda_2$ beam varies ($\lesssim$1\%) between the two polarization cases. We correct the ratio $S_{\parallel}/S_{\perp}$ to compensate for this effect, and estimate that the uncertainty in the average corrected power is $0.1-0.3$\%.

We rotate the HWP manually, and estimate the uncertainty in the orientation of the HWP as $\pm0.25^{\circ}$.  
We calculate that this introduces an uncertainty in $S_{\parallel}/S_{\perp}$ of $< 0.05\%$, and we apply a correction of the same magnitude to compensate.

A static magnetic field at the location of the cell (measured to be $\sim$0.5 Gauss due primarily to the Earth and the optical table) will cause a Zeeman splitting of the different magnetic components of the transition, which could cause an effective broadening of the transition.  For the parallel polarization case, only $\Delta m = 0$ transitions are allowed. Since we are driving only $\Delta F = 0$ transitions and the Land\'{e} $g$-factors are the same for the initial $6s\ ^2S_{1/2}$ and upper $7s\ ^2S_{1/2}$ state, the transition frequencies are unaffected.  For perpendicular polarization, however, $m$ does change ($\Delta m = \pm 1$), and so the transition frequency is affected by the magnetic field.  We model this as an effective broadening $\Delta \nu_Z$ of the homogeneous linewidth, and estimate the impact as a slight decrease in $S_{\perp}$ of magnitude $\sim \Delta \nu_Z / \Delta \nu_D \sim 0.2$\%.  To correct for this, we reduce each data point $S_{\parallel}/S_{\perp}$ by 0.1\%, and assign an uncertainty for this correction of 0.1\%.

The splitting of hyperfine levels of the $6p\ ^2P_{1/2}$ and $6p\ ^2P_{3/2}$ can affect the theory curves at small detunings.  We have analyzed the magnitude of this effect numerically, and find that for the range of detunings used for the measurements, the influence of the effect is much smaller than the experimental uncertainties.

We fit the spectral peaks with a Gaussian lineshape function in order to determine the peak amplitude of the fluorescence.  While a Voigt function, which is a convolution of the Lorentzian natural lineshape of width $\Delta \nu_n \sim 3.3$ MHz with the Gaussian inhomogeneous lineshape of width $\Delta \nu_D \sim 170$ MHz, would be more precise, $S_{\parallel}$ and $S_{\perp}$ are affected similarly, and the impact on the ratio $S_{\parallel}/S_{\perp}$ is minimal.  

Saturation of the two-photon transition rate can be a problem if laser intensities are too large.  We check for this by looking for any intensity dependence in the ratio $S_{\parallel}/S_{\perp}$.  We observe no such dependence at the level of our measurement precision. This is consistent with our estimate of the maximum two-photon transition rate per atom of $10^3$ s$^{-1}$, based upon the measured signal size, the PMT gain, and the estimated collection efficiency of the fluorescence detection.  Since this excitation rate is such a small fraction of the decay rate $\Gamma_{7s} = \tau_{7s}^{-1}$ of the $7s$ state, saturation effects are minimal. This lack of intensity dependence also rules out any significant effect of redistribution of the cesium ground state population by the lasers.  
\begin{table}[b!]
    \begin{tabular}{cccc}
        \hline
   $n$  &  $E_{np_{j}}$ (cm$^{-1}$)$^a$  & \rule{0in}{0.15in} $\langle 7s_{1/2} || r || np_{j}  \rangle $    	&  $\langle np_{j} || r || 6s_{1/2} \rangle $    \\  \hline \hline
   \multicolumn{2}{c}{$j=1/2$} & \rule{0in}{0.2in} & \\
   $6$  & $11178.268$ & -- 				 & $4.5057 \ (16)^b$ \\ 
   $7$  & $21765.348$ & $10.31 \ (4)^c$  & $0.2764 \ (4)^d$ \\ 
   $8$  & $25708.835$ & $0.914 \ (27)^c$ & $0.092 \ (10)^c$ \\ 
   $9$  & $27636.997$ & $0.349 \ (10)^c$ & $0.043 \ (7)^c$ \\ 
   $10$ & $28726.812$ & $0.191 \ (6)^c$  & $0.025 \ (5)^c$ \\ 
   $11$ & $29403.423$ & $0.125 \ (4)^c$  & $0.016 \ (4)^c$ \\ 
    \multicolumn{2}{c}{$j=3/2$} & \rule{0in}{0.2in} & \\
   $6$  & $11732.307$ & --            	 & $-6.3398 \ (22)^b$ \\ 
   $7$  & $21946.397$ & $14.32 \ (6)^c$  & $-0.5735 \ (7)^d$ \\ 
   $8$  & $25791.508$ & $1.620 \ (35)^c$ & $-0.232 \ (14)^c$ \\ 
   $9$  & $27681.678$ & $0.680 \ (14)^c$ & $-0.130 \ (10)^c$ \\ 
   $10$ & $28753.677$ & $0.396 \ (9)^c$  & $-0.086 \ (7)^c$ \\ 
   $11$ & $29420.824$ & $0.270 \ (7)^c$  & $-0.063 \ (6)^c$ \\ 
   \hline 
    \end{tabular}
    \caption{State energies and electric dipole $E1$ transition moments $\langle 7s_{1/2} || r || np_{j}  \rangle $ and $\langle np_{j} || r || 6s_{1/2} \rangle $ used to determine $R$.  Transition moments are given in terms of $a_0$.  $^a$State energies as found in NIST tables~\cite{kramida2016NIST_ASD}.  $^b$Weighted average of several independent determinations from Refs.~\cite{YoungHSPTWL94, RafacTLB99, DereviankoP02, AminiG03, BouloufaCD07, ZhangMWWXJ13,PattersonSEGBSK15,GregoireHHTC15}.  $^c$Ref.~\cite{SafronovaSC16}, including the Supplemental Information.  $^d$Ref.~\cite{DamitzTPTE18a}.}
    \label{table:E1moments}
\end{table}
We also considered any possible effects of radiation trapping (absorption and re-emission of 852 nm fluorescence photons before they can escape the vapor cell) on the measurement by measuring $S_{\parallel}/S_{\perp}$ at different vapor cell densities.  Since our measurement does not depend on timing of photon arrivals (as would be the case for a time-resolved lifetime measurement, for example), and since the signals $S_{\parallel}$ and $S_{\perp}$ would be affected similarly, it is difficult to identify a means by which radiation trapping affects the measurement of $S_{\parallel}/S_{\perp}$.  This is supported by our search for a dependence of this ratio on the vapor density in the cell, which had a negative result.

\section{Results}
\subsection{The ratio $R $}
We fit Eqs.~(\ref{eq:ratio44}, \ref{eq:ratio33} -- \ref{eq:beta_R}) to the data shown in Fig.~\ref{fig:ratiovsdetuning}, using just a single fitting parameter $R$, to determine the least squares fit value for this ratio of moments. In this fit, we use the lifetime $\tau_{7s}$~\cite{TohJGQSCWE18} as a constraint on the elements $\langle 7s_{1/2} || r || 6p_{3/2}  \rangle $ and $\langle 7s_{1/2} || r || 6p_{1/2}  \rangle $.  In order to evaluate $P$ and $Q$ of Eqs.~(\ref{eqn:sumP}, \ref{eqn:sumQ}), we use the state energies $E_{np_{j}}$ and the $E1$ transition moments for the $6s \ ^2S_{1/2} \rightarrow np \ ^2P_{J}$ and $7s \ ^2S_{1/2} \rightarrow np \ ^2P_{J}$, where $ n \le 11$ and $j = J = 1/2$ or $3/2$, listed in Table~\ref{table:E1moments}.
The state energies in this table come from Ref.~\cite{kramida2016NIST_ASD}.  The matrix elements come from a variety of experimental~\cite{YoungHSPTWL94, RafacTLB99, DereviankoP02, AminiG03, BouloufaCD07, ZhangMWWXJ13,PattersonSEGBSK15,GregoireHHTC15,DamitzTPTE18a} and theoretical~\cite{SafronovaSC16} works. 

We evaluate $\chi^2$, the sum of the squared deviations between data and best fit, each normalized by the uncertainty of the data point, to determine the uncertainty in R.  The reduced $\chi_r^2$ for this fit is  
$1.67$, indicating that some small additional errors are present in our measurement.  We increase our statistical error by $\sqrt{1.67}$ to accommodate these, and report a statistical error of $0.0016$, or $\sim 0.1$\%. 

Our result of the ratio $R$ can vary with the values of matrix elements used (shown in Table \ref{table:E1moments}) for curve fitting. We vary the values of the $ms \ ^2S_{1/2} \rightarrow np \ ^2P_{J}$ matrix elements used for fitting by their uncertainties, and found that most of them affect $R$ negligibly ($\pm 0.0001$). For the $n>6$ terms, this is reasonable since $P$ and $Q$ amount to only $\sim$1\% of the terms $\tilde{\alpha}$ and $\tilde{\beta}$, respectively. The uncertainty in the $6s \ ^2S_{1/2} \rightarrow 6p \ ^2P_{J}$ matrix elements resulted in the largest difference, a change in $R$ of $\pm 0.0006~(0.04\%)$. Adding this error in quadrature with our statistical error, our final result is $R = 1.5272 \ (17)$. 

\begin{table*}
  \begin{tabular}{lccc}
    \hline
      \multicolumn{1}{c} {Group}   		& Ratio $|R|$ & \hspace{0.02in} $| \langle 7s_{1/2} || r || 6p_{1/2} \rangle |$ \hspace{0.02in} & \hspace{0.02in} $| \langle 7s_{1/2} || r || 6p_{3/2} \rangle |$ \hspace{0.02in} \\ 
      \hline \hline 
  {\underline{\emph{Experimental}}} 		&    &    & \\
	 This work   							 & $1.5272 \ (17)$ & $4.249 \ (4)$ & $ 6.489 \ (5)$ \\
  & & \\
  \multicolumn{1}{l}{\underline{\emph{Theoretical}}} & & \\
     Dzuba {\it et al.}, 1989~\cite{DzubaFKS89}			& 1.530 & $4.253$ 	& $ 6.507$ \\
     Blundell {\it et al.}, 1991~\cite{BlundellJS91}  	& 1.526 & $4.228$ 	& $ 6.451$ \\
     Blundell {\it et al.}, 1992~\cite{BlundellSJ92}  	& 1.527 & $4.236$ 	& $ 6.470$ \\
     Safronova {\it et al.}, 1999~\cite{SafronovaJD99}	& 1.527 & $4.243$ 	& $ 6.479$ \\
     Dzuba {\it et al.}, 2001~\cite{DzubaFG01}  		& 1.526 & $4.255$  	& $ 6.495$  \\
     Porsev {\it et al.}, 2010~\cite{PorsevBD10}  		& --  	& $4.245$ 	& --  \\
     Present, Safronova {\it et al.}, 2016~\cite{SafronovaSC16}~~~~~  & 1.5270~(27)  &  $4.243 \ (11)$  &  $6.480 \ (19)$  \\
   \hline
  \end{tabular}
  \caption{Experimental and theoretical results for the ratio and absolute values of reduced dipole matrix elements for the cesium $6p \ ^2P_{J} \rightarrow  7s \ ^2S_{1/2} $ transitions.  We compute the ratio $R$ from the values of $\langle 7s_{1/2} || r || 6p_{1/2} \rangle$ and $\langle 7s_{1/2} || r || 6p_{3/2} \rangle$ reported in Refs.~\cite{BlundellJS91,BlundellSJ92,DzubaFKS89,SafronovaJD99,DzubaFG01,SafronovaSC16}. }
  \label{table:ResultCompRME}
\end{table*}

We use the lowest-order Dirac-Hartree-Fock (DHF) calculations to determine signs of all necessary matrix elements. We note that only relative signs are definite rather than the absolute signs. 
In the usual convention where the signs of the $\langle 6s_{1/2} || r || 6p_{j} \rangle$ matrix elements are positive, signs of $\langle 6s_{1/2} || r || np_{j} \rangle$ and $\langle 7s_{1/2} || r || np_{j} \rangle$ are positive, with the exception of the $\langle 7s_{1/2} || r || 6p_{j} \rangle$ matrix elements, which are negative. 
The signs of the $ns-n^{\prime}p_{j}$ and $n^{\prime}p_{j} -ns$ matrix elements are the same for $j=1/2$ and opposite for $j=3/2$.

In Table \ref{table:ResultCompRME}, we compare the measured result for $R$ with several theoretical calculations of this ratio.  We observe very close agreement between these results.  We are unaware of any prior experimental measurements of this ratio $R$.

Finally, we comment that our analysis based on a least-squares fit of $S_{\parallel}/S_{\perp}$ vs.~$\Delta$ differs from that used in Ref.~\cite{SieradzanHS04}, who defined  a linear polarization degree
\begin{equation}
  P_L = \frac{S_{\parallel} - S_{\perp}}{S_{\parallel} + S_{\perp}},
\end{equation}
and fit their data to this form to determine $R$.  These two analysis techniques likely place different weights to the various data points.  For comparison, we evaluated $R$ using this parameter as well, and find $R_{PL} = 1.5273 \ (17)$.  This is essentially the same result as we report in Table \ref{table:ResultCompRME}.

\begin{table}[b!]
  \begin{tabular}{lcccccc}
    \hline 
      \multicolumn{1}{c} {}   		&
      \multicolumn{1}{c} {DHF} &
        \multicolumn{1}{c} {SD} &
           \multicolumn{1}{c} {SD$_{\textit{sc}}$} &
           \multicolumn{1}{c} {SDpT} &
           \multicolumn{1}{c} {SDpT$_{\textit{sc}}$} &
         \multicolumn{1}{c} {Final} \\
         \hline \hline
$7s - 6p_{1/2}$ &	4.4177 & 4.2006 &	4.2434 &	4.2325 &	4.2313 & 4.243(11)\\
$7s - 6p_{3/2}$	&   6.6729 & 6.4258 &	6.4795 &	6.4608 &	6.4658 & 6.480(19)\\
$R$	            &   1.5105 & 1.5297 &	1.5270 &	1.5265 &	1.5281 & 1.5270(27)\\
\hline
  \end{tabular}
  \caption{The absolute values of the $7s - 6p_j$ reduced  dipole matrix elements (in $a_0$) and their ratio calculated in different approximations (see text for explanation).}
  \label{tabTh}
\end{table}
The results of several linearized coupled-cluster (LCC) \cite{SafJoh08,SafronovaSC16} calculations of the $7s-6p_j$ matrix elements and their ratio R are given in Table~\ref{tabTh}, with lowest order DHF values listed to show the effect of electronic correlations. \textit{Ab initio} LCC results obtained by taking into account single and double (SD) excitations of the lowest-order wave function are listed in the column labeled ``SD.'' The effect of partial triple excitations is accounted for in the SDpT calculations. The scaled SD and SDpT values are given in the corresponding columns. Following  Ref.~\cite{SafronovaSC16} and references therein, the SD scaled data are taken as final, based on the dominance of single-excitation valence terms, known cancellations of the triple contributions, and numerous comparisons with other experiments in many systems. 
The uncertainties in the values of matrix elements are determined as the maximum difference of the 
final and two other most precise results, \textit{ab initio} and scaled SDpT values. The uncertainty in the ratio is determined as the maximum difference of the 
final and all other LCC values. The issue of the accuracy of the ratio is the long-standing question - does scaling adversely affect the ratio precision? The present experiment provides a benchmark comparison to address this question. The final theory value is well within 1 (experimental) $\sigma$ from the central experimental value while  the SD value is approximately $2\sigma$ away - so further inclusion of the correlations via the SDpT method or scaling improved the agreement with experiment.

\subsection{Absolute Matrix Elements }
In this section, we combine the ratio of matrix elements $R = \langle 7s_{1/2} || r || 6p_{3/2} \rangle / \langle 7s_{1/2} || r || 6p_{1/2} \rangle = 1.5272 \ (17)$ with the lifetime result that we reported previously~\cite{TohJGQSCWE18} of the cesium $7s \ ^2S_{1/2}$ state, $\tau_{7s} = 48.28 \ (7)$~ns.  This lifetime can be written in terms of the matrix elements as 
  \begin{equation}
     \frac { 1 }{ { \tau_{7s}  } } = \sum_{j = 1/2, 3/2}     \frac { 4 }{ 3 }  \frac { { \omega_{j}}^{ 3 } }{ { c }^{ 2 } } \alpha \frac { { \left| \langle 7s||r||6p_{j} \rangle \right|  }^{ 2 } }{ 2j^{\prime}+1 } .
   \end{equation}
In this equation, $j^{\prime} = 1/2$ is the electronic angular momentum of the $7s \ ^2S_{1/2}$ state, $\omega_j$ are the transition frequencies for the $7s \ ^2S_{1/2} \rightarrow 6p \ ^2P_{J}$ transitions (where $j = J$), and $\alpha$ is the fine structure constant.  
The results of these two works combined uniquely determine the individual matrix elements $\langle 7s_{1/2} || r || 6p_{3/2} \rangle = -6.489 \ (5)$ and $\langle 7s_{1/2} || r || 6p_{1/2} \rangle = -4.249 \ (4)$. These results are in very good agreement with theoretical calculations, as we present in Table \ref{table:ResultCompRME}.

\section{Conclusion}
We have described our laboratory measurement of the ratio $R = \langle 7s_{1/2} || r || 6p_{3/2} \rangle / \langle 7s_{1/2} || r || 6p_{1/2} \rangle = 1.5272 \ (17)$, whose precision is $\sim$0.11\%.  We determine this ratio through observations of the two-color two-photon absorption rate to the $7s \ ^2S_{1/2}$ state with two different polarization cases over a broad range of detunings of the laser frequency from the D$_2$ resonance frequency.  Combined with an earlier lifetime measurement~\cite{TohJGQSCWE18} for the $7s \ ^2S_{1/2}$ state, we present experimental determinations of the individual matrix elements $\langle 7s_{1/2} || r || 6p_{3/2} \rangle $ and $ \langle 7s_{1/2} || r || 6p_{1/2} \rangle $, with uncertainty of $<$0.1\%.  These measurements are in very good agreement with theoretical calculations of these moments.  

These measurements bring to near completion a series of precision determinations of each of the matrix elements $\langle ns_{1/2} || r || mp_{j} \rangle$ for $m, n = 6$ or $7$.  We will report the final missing element $\langle 7p_{j} || r || 6s_{1/2} \rangle$ shortly in a separate publication. 

This material is based upon work supported by the National Science Foundation under Grant Numbers PHY-1607603, PHY-1460899 and PHY-1620687.

\bibliography{biblio}

\end{document}